\documentclass[12pt,usenatbib]{mn2e}
\usepackage{graphicx}
\usepackage{epsf}
\usepackage{color}
\usepackage{rotating}
\def\lsim{\mathrel{\lower0.6ex\hbox{$\buildrel {\textstyle <}
 \over {\scriptstyle \sim}$}}}
\def\gsim{\mathrel{\lower0.6ex\hbox{$\buildrel {\textstyle >}
 \over {\scriptstyle \sim}$}}}

\def\eone{$\hat{e}_{1}$}
\def\etwo{$\hat{e}_{2}$}
\def\ethree{$\hat{e}_{3}$}
\def\lamone{$\lambda_{1}$}
\def\lamtwo{$\lambda_{2}$}
\def\lamthree{$\lambda_{3}$}

\begin{document}

\title[The velocity shear tensor]{The velocity shear tensor: tracer of halo alignment}
\author[Libeskind et al.] 
{Noam I Libeskind$^{1,2}$, Yehuda Hoffman$^3$, Jaime Forero-Romero$^{4,5}$, Stefan Gottl\"ober$^1$, \newauthor Alexander Knebe$^6$, Matthias Steinmetz$^1$, Anatoly Klypin$^{7}$\\
 $^1$Leibniz-Institute f\"ur Astrophysik Potsdam (AIP), An der Sternwarte 16, D-14482 Potsdam, Germany\\
 $^2$Kavli Institute for Theoretical Physics, University of California Santa Barbara CA 93106-4030, USA\\
 $^3$Racah Institute of Physics, Hebrew University, Jerusalem 91904, Israel\\
  $^4$Department of Astronomy, University of California, Berkeley, CA 94720-3411, USA\\
$^{5}$Departamento de Fisica, Universidad de los Andes, Cra. 1 No. 18A-10, Edificio Ip, Bogota, Colombia\\
  $^6$Grupo de Astrofisica, Departamento de Fisica Teorica, Modulo C-8, Universidad Aut\'onoma de Madrid, Cantoblanco E-280049, Spain\\
$^7$Astronomy Department, New Mexico State University, Las Cruces, NM 88003, USA\\
  }
\date{Accepted --- . Received ---; in original form ---}
\pagerange{\pageref{firstpage}--\pageref{lastpage}} \pubyear{2002}
\maketitle

 \begin{abstract}
The alignment of dark matter (DM) halos and the surrounding large scale structure (LSS) is examined in the context of the cosmic web. Halo spin, shape and the orbital angular momentum of subhaloes is investigated relative to the LSS using the eigenvectors of the velocity shear tensor evaluated on a grid with a scale of 1 Mpc/h, deep within the non-linear regime. Knots, filaments, sheets and voids are associated with regions that are collapsing along 3, 2, 1 or 0 principal directions simultaneously. Each halo is tagged with a web classification (i.e. knot halo, filament halo, etc) according to the nature of the collapse at the haloÕs position. The full distribution of shear eigenvalues is found to be substantially different from that tagged to haloes, indicating that the observed velocity shear is significantly biased. We find that larger mass haloes live in regions where the shear is more isotropic, namely the expansion or collapse is more spherical. A correlation is found between the haloÕs shape and the eigenvectors of the shear tensor, with the longest (shortest) axis of the haloÕs shape being aligned with the slowest (fastest) collapsing eigenvector. This correlation is web independent, suggesting that the velocity shear is a fundamental tracer of the halo alignment. A similar result is found for the alignment of halo spin with the cosmic web. It has been shown that high mass haloes exhibit a spin flip with respect to the LSS: we find the mass at which this spin flip occurs is web dependent and not universal as suggested previously. Although weaker than haloes, subhalo orbits too exhibit an alignment with the LSS, providing a possible insight into the highly correlated co-rotation of the Milky WayÕs satellite system. The present study suggests that the velocity shear tensor constitutes the natural framework for studying the directional properties of the non-linear LSS and of halos and galaxies.
\end{abstract}

\section{Introduction}
\label{introduction}

 \begin{figure*}
 \includegraphics[width=40pc]{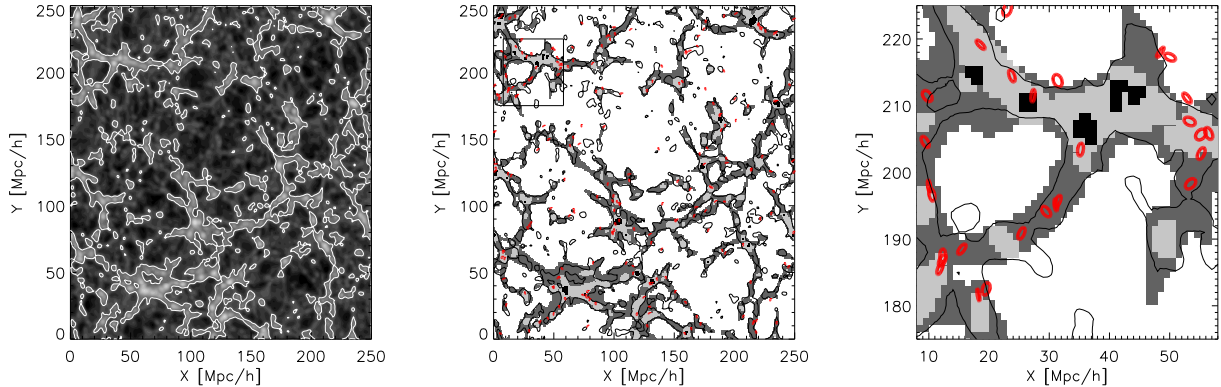}
 \caption{The cosmic web in a 0.97~Mpc$/h$ (one grid cell) slice. \textit{Left panel:} the DM density distribution, with contours denoting regions that are over dense with respect to the mean. \textit{Middle panel:} The cosmic web in the same slice with the most massive haloes over plotted in red. The hierarchy of cosmic web types is visible as knots (black) are embedded in filaments  (light gray) which are embedded in sheets (dark gray) which surround voids (white). Note that due to the thinness of this slice, sheets appear one dimensional. We over plot regions above the mean over density as empty contours. \textit{Right panel:} A zoom into a 50~Mpc$/h$ sub region (denoted in the middle panel by the small square) showing each halo as an ellipse whose major and minor axes are defined by the inertia tensor. Again, regions more over dense than the mean are shown with the thin black contours. The haloes are aligned with large scale structure as shown in Fig.~\ref{fig:halo-shape}. Note that the size of each halo is not representative of the halo's virial radius.}
 \label{fig:cosmicweb}
 \end{figure*}

According to the current cosmological paradigm, the early Universe was seeded with small density perturbations that peppered an otherwise homogeneous matter distribution. As the Universe expanded, these peaks in the density field grew due to gravitational instability at the expense of under dense regions producing a highly complex network of voids, walls, filaments and knots, known as the cosmic web. The first attempt to quantify the growth of perturbations and thus the cosmic web is attributed to  \cite{1970A&A.....5...84Z} who's famous ``pancake'' solution forms the basis of how we think about the cosmic web. Further work \citep[by, e.g.][and references therein]{1996Natur.380..603B,2004MNRAS.353..162S,2006ApJ...645..783S} began to examine in great detail the nature of how the cosmic web formed and its defining properties.

This theoretical view is corroborated by mapping the distribution of galaxies on the largest scales \citep[by e.g. the 2dFGRS and SDSS,][]{2001MNRAS.328.1039C,2000AJ....120.1579Y} which quite clearly defines the cosmic web \citep{2004MNRAS.352..939E}. Many studies have examined web attributes in detail focusing on knots \citep{1997A&AS..123..119E}, filaments \citep{1996Natur.380..580C,2004ApJ...606...25B}, sheets \citep{2011ApJ...736...51E} and voids \citep{2011AJ....141....4K,2012MNRAS.421..926P,2008MNRAS.385..867C}. Recently, \cite{2011MNRAS.417.1303M} used the data from the SDSS to reconstruct the cosmic density field, including the cosmic web.

Identifying the cosmic web in numerical simulations poses a number of interesting challenges. The simplest techniques \citep[e.g][]{2005ApJ...634...51A,2008MNRAS.383.1655S} define the web purely geometrically, often only in terms of the density field smoothed over a few Megaparsecs. \cite{Hahnetal2007a} and \cite{Forero-Romero2009} developed the concept using a dynamical stability criterion (the Hessian of the potential)  which in many ways is more fundamental than the density field since it reflects the dynamics of matter. Since the tidal web of \cite{Hahnetal2007a} is in principle only applicable to regions where the tidal tensor is representative of the large scale dynamics, it is incapable of identifying the finer web seen in high resolution numerical simulations. \cite{2012arXiv1201.3367H} extended these ideas by suggesting that the velocity shear (so-called ``V-web'') is a better representation of the cosmic web on linear and sub-linear scales (the velocity shear tensor converges to the tidal tensor in the linear regime).

 \begin{figure*}
 \includegraphics[width=40pc]{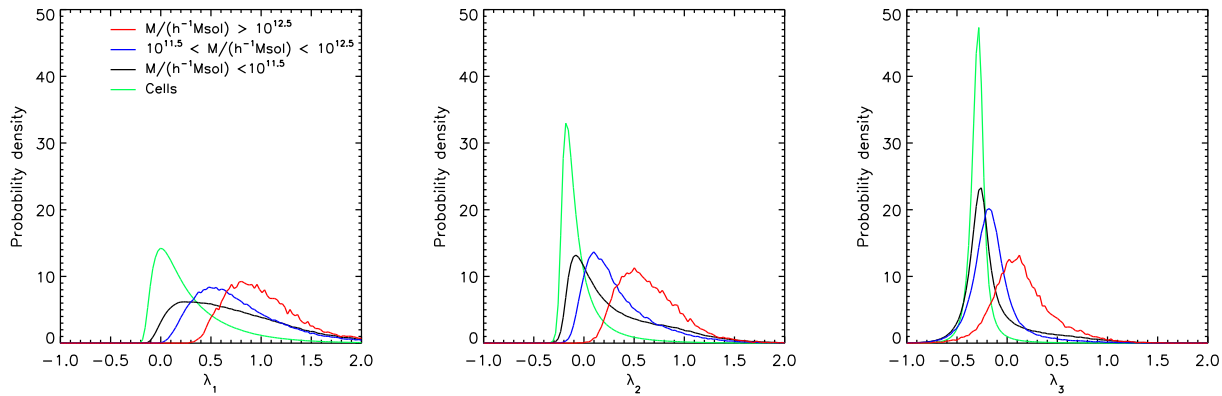}
 \caption{The probability density distribution for the largest (left), intermediate (middle) and smallest (right) eigenvalues of the velocity shear tensor. The distribution of eigenvalues for all cells is shown in green. We also present the distributions of eigenvalues assigned to low (black), intermediate (blue) and high (red) mass haloes.  Each distribution is normalized to the number of objects it contains. The noise in the high mass (red) curve, is due to poorer number statistics in that mass bin. Note that the distribution of eigenvalues assigned to haloes are significantly different from those of the full cosmological volume: just as haloes trace the highest peaks in the density field, they are also a biased tracer of velocity shear.}
 \label{fig:lambda-distr}
 \end{figure*}

While the cosmic web is interesting in and of itself, its characteristics affect the haloes and galaxies that inhabit it. For example, tidal torque theory \citep{1969ApJ...155..393P,1984ApJ...286...38W}, the theory by which galaxies acquire angular momentum, suggests that the large scale tidal force field is responsible for both the formation of the cosmic web and the spin of galaxies. A correlation between galaxy spin and web environment is thus expected. This correlation has been found in a few observational studies \citep{1982A&A...107..338B,2012arXiv1207.0068T}  that defined aspects of the cosmic web using various techniques. The recent work of \cite{2012arXiv1207.0068T} found a weak (but significant) trend for the spin axis of disc galaxies to be aligned with filaments, while that of ellipticals appears to be perpendicular to them. Yet these observational studies of environment are really in their infancy, since obtaining a three dimensional spin vector for each galaxy as well as a proper classification of the cosmic web is difficult (if not impossible when, for example, considering the angular momentum of elliptical galaxies).

A number of numerical studies \citep[e.g.][]{2006MNRAS.370.1422A,Aragon-Calvoetal2007,2007MNRAS.375..184B,2012arXiv1201.5794C} have examined the orientation of halo shape or angular momentum with respect to the cosmic web in just one or two given web types. For example \cite{2006MNRAS.370.1422A} looked at the alignment between halo shape and the main filament axis while
 \cite{2012arXiv1201.5794C} examined the orientation of halo spin for haloes found in filaments. \cite{Aragon-Calvoetal2007} looked at the orientation of angular momentum of filament and sheet haloes while \cite{2007MNRAS.375..184B} examined both spin and shape alignments in voids found in the Millenium simulation \citep{2006Natur.440.1137S}. The recent work of \cite{2012arXiv1201.6108T} also used the Millenium simulation to examine the evolution of low and high mass halo spin alignments within filaments.

The general consensus from most of these studies is two fold: the first is that halo shape correlations are stronger than spin correlations. Second is that haloes are oriented (at the $\sim 20\%$ level) with the cosmic web. Haloes tend to point along the spines of filaments, parallel to the plane of cosmic sheets, and perpendicular to the radial direction of voids. \cite{Aragon-Calvoetal2007}  and \cite{2012arXiv1201.5794C} found that the angular momentum axis of low mass haloes points along the filament axis while that of high mass haloes points perpendicular to it. The mass at which this spin flip is seen was reported to be $\sim1-5\times 10^{12}M_{\odot}$ and is possibly due to the fact that high mass haloes grow by mergers which preferentially come from the filament axis, while low mass haloes are ``wrapped up'' with the filament as it collapses. While clearly an attractive explanation, the critical mass of the alignment transition appears empirical and arbitrary since it has not been theoretically motivated. Moreover, \cite{2012arXiv1201.6108T} examined the evolution of the spin alignment with redshift and found a completely opposite effect: all haloes are oriented perpendicular to their filaments axis at high redshift and it is only low mass haloes which become parallel later on. Clearly the origin of the spin web alignment requires significant study.

\cite{2012MNRAS.421L.137L} used the V-web to examine the orientation of haloes and, for the first time, their subhaloes with respect to the cosmic web in simulations tuned to reproduce the local universe. They found that the V-web confirmed previous studies that showed that high (low) mass halos tend to spin perpendicular (parallel) to the defining filamentary axis. Due to its limited sample size it was unable to measure the transitional mass at which halo spin's flip. However, they found that subhalo orbits as well as halo spins are correlated with the large scale velocity shear. This result was surprising since subhaloes orbit deep in their parent potential, in a way shielded from the large scale velocity shear. That the cosmic web can still be reflected amidst subhalo orbits, is a testament to its importance in galaxy formation.

The finding that subhaloes reflect the cosmic web is suggested by studies that focus on the alignment of subhalo positions and velocities with respect to each other and their host \citep[e.g.]{Libeskindetal2005,Zentneretal2005,Vera-Ciroetal2011,Libeskindetal2011a}. Yet \cite{KroupaTheisBoily2005} and more recently \cite{2012arXiv1204.6039P} have argued that the satellites of the Milky Way can not be $\Lambda$CDM substructures since their anisotropic spatial and kinematic $z=0$ distribution is incompatible with simulations, which in a naive sense predict an isotropic spatial distribution with randomized orbits. Thus the alignment between substructures and their large scale structure is of great interest in the study of the Milky Way.

In this paper, we extend the work of \cite{2012MNRAS.421L.137L} and examine halo and satellite alignments with respect to the large scale structure as defined by the V-web. We take a detailed look at both the cosmic web itself, by studying the eigensystems of the diagonlized shear tensor, and the effect on haloes within it. Numerical methods and the algorithm used to dissect the cosmic web are explained in Section~\ref{sec:methods}. The main results with respect to the cosmic web and halo alignment are presented in Section~\ref{sec:results}. We conclude in Section.~\ref{sec:conclusion}.

\section{Methods}
\label{sec:methods}

\subsection{The Bolshoi Simulation}

In order to investigate correlations with the cosmic web, we use the Bolshoi \citep{2011ApJ...740..102K} $N$-body simulation. We refer the reader to \cite{2011ApJ...740..102K} for details on the simulation, and include just the salient points here. 
The run follows the evolution of matter in a large periodic box according the $\Lambda$CDM paradigm, employing cosmological parameters that are compatible with WMAP7 \citep{2011ApJS..192...14J} data, e.g. $h= 0.7$, $\Omega_{\Lambda}=0.73$, $\Omega_{\rm m}=0.27$, $\Omega_{b}= 0.046$, $n=0.95$ and $\sigma_{8}=0.82$. Dark matter (DM) is numerically simulated using an Adaptive-Refinement-Tree code \citep[ART,][]{1997ApJS..111...73K}. The simulation box has a side length of 250~$h^{-1}$~Mpc and is filled with 2048$^{3} \approx$ 8 billion particles, each with a mass of $\sim 1.3\times10^{8}~h^{-1}M_{\odot}$. The proper physical spatial resolution (smallest cell size) of the simulation is $1 h^{-1}$~kpc. In this paper we use only the simulations $z=0$ snapshot.

DM halos are identified using a Friends-of-Friends \citep[FOF][]{1985ApJ...292..371D} algorithm with $b=0.17$. FOF links particles closer than $b$ times the mean inter particle separation, into objects referred to as haloes. In effect, the free parameter $b$ defines  isodensity contours in the matter distribution. The choice of $b=0.17$ returns haloes whose mean over-density is roughly equal to that of virialized objects in the CDM cosmology. For each halo, internal properties such as mass and angular momentum (spin), are calculated. The shape of each halo is defined by diagonalizing the reduced \citep{1983MNRAS.202.1159G,2005ApJ...627..647B} inertia tensor $I_{ij}$, constructed by summing over $n$ particles in the halo:
\begin{equation}
I_{ij}=\sum_{n}\frac{x_{i,n}x_{j,n}}{r_{n}^{2}}
\label{eq:inertia}
\end{equation}
where $r_{n}=(x_{n}^{2}+y_{n}^{2}+z_{n}^{2})$ is the distance to each $n$th particle. The reduced inertia tensor is used instead of the standard one (namely without the $r^{2}_{n}$ term in the denominator) so as to lessen the effect of recently accreted large subhaloes in the outer parts of the halo. The three eigenvalues ($a$, $b$ and $c$) are sorted in increasing order such that $a>b>c$. The halo's principle axes ($\hat{a}$, $\hat{b}$ and $\hat{c}$) are associated to the eigenvectors that correspond to each eigenvalue. An arbitrary mass cut of $10^{9.5}h^{-1}M_{\odot}$ is applied to the halo catalogue such that very small haloes are excluded from the analysis.

Substructures are identified using the Bound-Density-Maxima  \textsc{BDM} technique \citep{1997astro.ph.12217K}.  \textsc{BDM} finds local maxima in the density field, removes unbound particles, and computes the relevant halo properties (such as mass) within a truncated radius. We impose a 30 particle limit on the BDM subhaloes, ignoring smaller clumps\footnote{The Bolshoi simulation (halos as well as the DM particle positions and velocities) can be downloaded from the MultiDark database (http://www.multidark.org). The databases as well as the FOF and BDM halo finder used here are described in \cite{2011arXiv1109.0003R}.}. Note that the BDM catalogues are only used to identify the orbital angular momentum of subhaloes.

Throughout this paper, haloes are grouped according to mass into three bins, a low mass bin ($M_{\rm vir} <10^{11.5}h^{-1}M_{\odot}$, colored black), an intermediate ``Milky Way'' mass bin  ($10^{11.5}h^{-1}M_{\odot}<M_{\rm vir} <10^{12.5}h^{-1}M_{\odot}$, colored blue), and a high mass bin ($M_{\rm vir} >10^{12.5}h^{-1}M_{\odot}$, colored red).

 \begin{figure}
 \includegraphics[width=20pc]{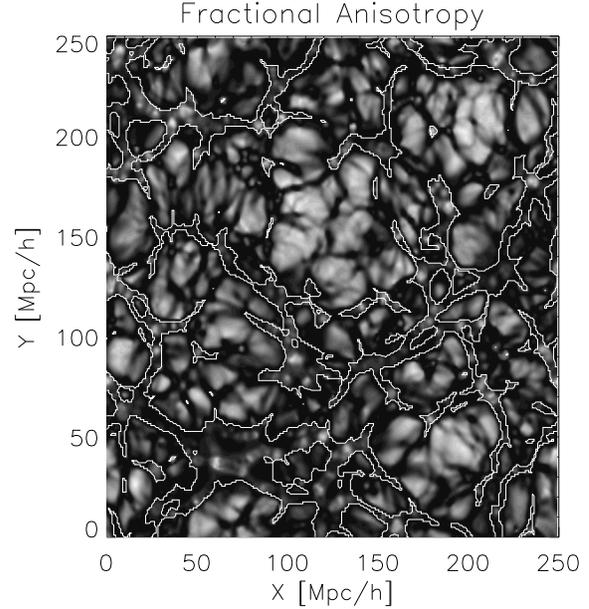}
 \caption{The Fractional Anisotropy (\textit{FA}), as defined in equation~\ref{eq:fa} is a measure of shear anisotropy. The same slice of the simulation shown in Fig.~\ref{fig:cosmicweb}. The thin white lines separate voids from sheets, filaments and knots. The continuous black and white regions denote low and high values of \textit{FA}, ranging from zero to unity. Note how the higher values of \textit{FA} are found abutting the contours and snaking through the voids, revealing the inner structure of voids.}
 \label{fig:cosmicweb2}
 \end{figure}

 \begin{figure}
 \includegraphics[width=20pc]{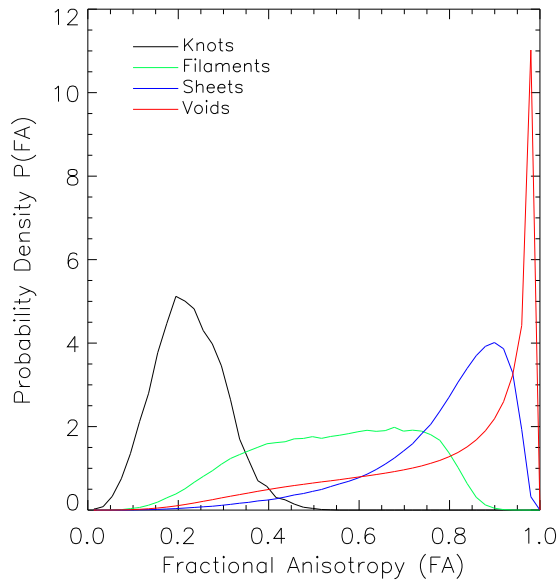}
 \caption{The Fractional Anisotropy \textit{FA} is a measure of shear anisotropy. Since each halo can be assigned a web type based on the local velocity shear calculated at its location, we divide our halo sample into four web classes, plotting the probability density distribution of \textit{FA} for haloes resident in knots (black), filaments (green), sheets (blue) and voids (red). Each distribution is normalized by the number of objects in it. \textit{FA}=0 indicates uniform expansion or collapse. Since the peaks and widths of these distributions depend on web type, we infer that each web type behaves significantly different: knot collapse is not simply void expansion played in reverse.} 
 \label{fig:fa-distr}
 \end{figure}

\subsection{The Cosmic Web}
In order to quantify the cosmic web, the velocity shear tensor of the particle distribution is computed \citep[see][for a detailed analysis of this technique]{2012arXiv1201.3367H}. First, a $256^{3}$ grid based density and velocity field is constructed using the ``clouds in cells'' (CIC) technique, resulting in a smoothed density and velocity distribution. The CIC'd velocity field is then Fourier Transformed into $k$-space and smoothed with a Gaussian of kernel width equal to one grid cell ($0.97h^{-1}$~Mpc) in order to wash out artificial effects introduced by the preferential axes of the Cartesian grid. Using the Fourier Transform of the velocity field the normalized shear tensor is calculated as:
\begin{equation}
\Sigma_{\alpha\beta} = -\frac{1}{2H_{0}}\left(\frac{\partial v_{\alpha}}{\partial r_{\beta}}+\frac{\partial v_{\beta}}{\partial r_{\alpha}}\right)
\end{equation}
where $\alpha, \beta =x,y,z$. Similar to the inertia tensor, the eigenvalues (\lamone, \lamtwo~and \lamthree)  and corresponding eigenvectors (\eone, \etwo~and \ethree)  of the shear tensor are obtained at each grid cell. Following convention, the eigenvalues are ordered such that $\lambda_{1}>\lambda_{2}>\lambda_{3}$ and a web classification scheme based on how many eigenvalues are above an arbitrary threshold is carried out. If none, one, two or three eigenvalues are above this threshold, the grid cell is classified as belonging to a void, sheet, filament or knot. 

Finding the threshold which best returns the visual impression of a cosmic web is not self-evident, as it depends on the numerical parameters used. \cite{2012arXiv1201.3367H} and \cite{2012MNRAS.421L.137L}, found a threshold of $\lambda_{\rm th}=0.44$ best reproduced the visual impression of the cosmic web. This choice of threshold implies that when the expansion (or contraction) is very weak, it is considered contraction (or expansion). This threshold is used in this work as well, following the same visual justification.  We defer to a future study the effect of varying the threshold on the cosmic web. 

The cosmic web of the simulation is presented in Fig~\ref{fig:cosmicweb}, where we show a thin slice through the computational box. The density field is depicted in the left panel, with contours enclosing regions that are above the mean density. This can be compared with the cosmic web of the same slice shown in the middle panel of Fig,~\ref{fig:cosmicweb} (with haloes over plotted in red). The reader will note that more than 80\% of the regions of the simulation that are above the mean over density are designated as sheets, filaments and knots by the web classifier. The right most panel of Fig,~\ref{fig:cosmicweb} shows a zoom on a random sub-volume in the box, chosen to highlight the alignment of haloes with the large scale structure, as described in section~\ref{section:halo-shape}.

Haloes (and their subhaloes) are connected to a local velocity shear tensor by simply identifying the CIC grid cell in which the halo sits. Properties of the cosmic web (as calculated by the local shear tensor) in that cell are then associated to each halo within it - specifically, each halo is assigned environmental attributes, such as shear eigenvalues, eigenvectors and web type. In Section~\ref{section:eigenratio} we associate the uniformity of the velocity shear to haloes as well.

The discrete sampling of the cosmic density field is an attempt to numerically represent the underlying fluid. Naturally, any technique which attempts to redraw a continuous field in discrete terms will undoubtedly be plagued by artifacts introduced by the sampling itself. When examining the cosmic density field, these problems are acute since the density field spans many orders of magnitude and is considerably heterogenous, while the grid is regular. Ideally the cell size and smoothing kernel used would be adaptive and varied based on the local properties of the underlying density field. In the absence of an adaptive grid, the CIC and the Gaussian kernel chosen for the smoothing set a physical scale on the objects which can be examined. A grid size that is smaller than a halo's virial radius will probe the shear tensor interior to that halo, whereas a much coarser one will calculate the shear in a larger region, thereby washing out any inherent spin-web correlation. Thus the careful selection of an appropriate grid size and smoothing scale are crucial.

In this work we choose a 256$^{3}$ grid, which spans the 250Mpc$/h$ cosmological box, giving a cell size of $\approx 1$Mpc$/h$. All objects whose virial radius is smaller than this will be embedded in a cell: some will fit snugly, others more loosely. This grid size has been chosen such that all the objects in our halo catalogue can be reliably examined (i.e. a negligible number of haloes have $r_{\rm vir} \gsim 1$Mpc$/h$.) In practice this choice of grid puts an upper limit of $\sim2\times10^{14}M_{\odot}/h$ on the halo mass we may probe.

\section{Results}
\label{sec:results}
Before presenting results on the alignments of haloes with respect to the eigenvectors of the velocity shear tensor, we focus on understanding the distribution of the eigenvalues, \lamone, \lamtwo~and \lamthree. The range of the smallest and largest eigenvalues is essentially bounded by the value of $\lambda_{2}$, while the intermediate eigenvalue is constrained by the values of \lamone~and \lamthree. That is, 
\begin{eqnarray*}
\lambda_{1} &\in& [\lambda_{2},\infty)\\
\lambda_{2} &\in& [\lambda_{3},\lambda_{1}]\\
\lambda_{3} &\in& (-\infty,\lambda_{2}]
\end{eqnarray*}
 
The eigenvalue describes the strength of the expansion or collapse of matter along the corresponding eigenvector. For example, three large positive eigenvalues indicate strong collapse along all three axes. On the other hand in filaments and sheets where \lamone$>\lambda_{\rm th}$ and \lamthree$<\lambda_{\rm th}$ expansion along one axis is roughly equal to collapse along another when $|$\lamthree~$|\sim|$\lamone$|$.  Note that the trace of the shear tensor is
\begin{eqnarray}
\label{eq:trace}
{\rm\bf Tr}(\Sigma)&=&\lambda_{1}+\lambda_{2}+\lambda_{3}\\
	&=&-\nabla\cdot\vec{\rm V}\propto \delta
\end{eqnarray}
and is thus proportional to the over density $\delta$ in the linear regime.

We now turn to the distribution of velocity shear eigenvalues and show, in Fig.~\ref{fig:lambda-distr}, the normalized probability density distribution of \lamone, \lamtwo, and \lamthree. Since each halo can be assigned an eigen-system based on its location, both the full distributions of eigenvalues for all cells (green lines), as well as subsets of eigenvalues that are assigned to haloes binned by halo mass (black, blue and red lines) are shown. Note that the latter is identical to the distribution of eigenvalues weighted by halo number. Each distribution is normalized to unity.

The full distribution of $\lambda$'s (green curves) shows a number of interesting properties. Firstly, they are all well represented by log-normal distributions, whose median and width are eigenvalue dependent. Secondly, \lamthree~is not simply a negative reflection of  \lamone, as would be expected if expansion and collapse in different regions of space directly correlated with each other (i.e. if collapse looked like expansion in reverse). Instead the width of the distributions decrease with decreasing eigenvalue: \lamone~has a relatively wide width while \lamthree~is much narrower. This implies that expansion or collapse along \eone~may occur with varying strengths while the range of  the expansion or collapse strength along \ethree~is more tightly restricted.

The distribution of eigenvalues for those regions of space that contain haloes (black, blue and red curves in Fig.~\ref{fig:lambda-distr}) shows a number of interesting attributes as well.  Despite all being drawn from the same parent distribution (green curves), they are qualitatively very different. Firstly we note that none of the eigenvalue distributions (when binned by mass) resemble the parent distribution in terms of width or median. The low mass bin (black curve) peaks at a similar value to the parent distribution, probably owing to the fact that this mass bin spans the largest fraction of grid cells. Yet the peak of each distribution is clearly mass dependent. The more massive the set of haloes being examined, the greater the median of the distribution. Intuitively this is because the most massive haloes reside in regions of strong gravitational collapse that is, regions with larger values of $\lambda$.
 
\subsection{The Fractional Anisotropy: The distribution of shear sphericity}
\label{section:eigenratio}

The semi-infinite range of eigenvalues (described above), makes it difficult to characterize the uniformity of the collapse or expansion in the way that, for example, the inertia tensor may be characterized by its triaxiality. In their seminal work on Gaussian random fields,  \cite{1986ApJ...304...15B} defined the ``eccentricity''  $e$ of a density field (not a shear), as 
\begin{equation}
e=\frac{\lambda_{1}-\lambda_{3}}{2|\delta|} 
\label{eq:ecc}
\end{equation}
 where in the linear regime the dimensionless over density is defined as  the trace of the shear, i.e. equation~\ref{eq:trace}. This measure of eccentricity has been used by a number of authors \citep[e.g.][]{2008ApJ...688...78L} when examining velocity shears. That said, the original Bardeen et al. definition was invoked in a system with three positive eigenvalues (such as the inertia tensor) and not in the situation where the eigenvalues are bound by $(-\infty,\infty)$ as in the shear tensor.

The usage of the Bardeen et al. eccentricity is therefore highly problematic in our situation. In cases where $\lambda_{1}\sim-\lambda_{3}$, small values of \lamtwo~can cause the denominator of equation~\ref{eq:ecc} to tend to 0 making $e$ highly unstable. Also, the fact that $e$ is semi-infinite makes it more difficult to compare vastly different eccentricities.

In order to overcome these difficulties, the amount of anisotropy in the shear tensor can be characterized using the \textit{fractional anisotropy}, \textit{FA}
\begin{equation}
\label{eq:fa}
{\rm FA}=\frac{1}{\sqrt{3}}\sqrt{\frac{ (\lambda_{1}-\lambda_{3})^{2}+(\lambda_{2}-\lambda_{3})^{2}+(\lambda_{1}-\lambda_{2})^{2}}{\lambda_{1}^{2}+\lambda_{2}^{2}+\lambda_{3}^{2}}}
\end{equation}
which measures the fraction of the magnitude of the shear that is due to the anisotropy of the eigenvalues \citep{Basser1995}
\footnote{The concept of the \textit{Fractional Anisotropy} was developed in the field of Nuclear Magnetic Resonance by \cite{Basser1995} as a tool in brain imaging. The diffusivity of water molecules through cerebral tissue, one of the quantities measured by an \textit{NMR} scan, constitutes an identical process to the velocity shear. Thus both the cosmic velocity shear tensor and the diffusion tensor in brain scans can be quantified by the \textit{Fractional Anisotropy}.}. Note that the $1/\sqrt{3}$ is introduced to normalize the fraction to unity. This definition is employed to characterize the uniformity of the collapse or expansion.

The \textit{FA} measures the kinematical ``morphology'' of the velocity shear. It is superior to the eccentricity since it is forced to be between zero and unity. \textit{FA}=0 implies isotropic expansion or collapse in all three directions. In filaments and sheet volumes, \textit{FA}=1 indicates that collapse along \eone~is roughly the same strength as expansion along \ethree, while in knots and voids, \textit{FA}=1 indicates large anisotropies in the expansion or contraction.

In Fig.~\ref{fig:fa-distr} we show the distribution of the local velocity shear \textit{FA}. Since each web type represents a different number of positive eigenvalues we plot them seperately. A number of important points can be gleaned from this figure. 
\begin{itemize}
\item The collapse in knot environments is relatively isotropic; that is, the value of \textit{FA} measured in knots tends to be low. The strength of the velocity shear in knot environments thus displays a remarkable regularity.
\item The distribution of \textit{FA} in filaments and sheets is wide  - the shear that defines these web types takes on a myriad of kinematical morphologies. Whereas filaments peak at lower values of \textit{FA}, sheets peak at higher values.,
\item Void regions have high values of \textit{FA} indicating highly anisotropic expansion.
\end{itemize}

We thus conclude from Fig.~\ref{fig:fa-distr} that the cosmic web reflects a remarkable variety of shear anisotropies. Voids for example, are often not kinematically isotropic objects. Knots display significantly more uniformity than voids, although also span a wide range of sphericities. Filaments and sheets on the other hand, while displaying some similarities are still coloured by a variety of corpulence: some filaments are fat, some are thin, some sheets are thick some are narrow.

The \textit{FA} is shown graphically in Fig.~\ref{fig:cosmicweb2}. Since voids dominate the volume, the figure emphasizes the rich diversity in the kinematic ``morphology'' of voids. Although it appears that much of the void volume is represented by regions of high \textit{FA} (white areas in Fig.~\ref{fig:cosmicweb2}), void regions that abut sheets have lower \textit{FA} (black regions in Fig.~\ref{fig:cosmicweb2}). Interestingly this plot shows exactly why the \textit{FA} spans the full range of values in voids: deep inside voids, structures of low \textit{FA}  can be seen snaking through the expanse. The \textit{FA} is thus a measure that can be used to probe the inner morphological structure of voids similar to \cite{2012arXiv1203.0248A}. Recall that the \textit{FA} is insensitive to the \textit{sense} of the anisotropy and will return the same value for uniform expansion as for uniform contraction.

Armed with an understanding of the distribution of shear tensor anisotropy, we proceed to examine how haloes and their subhaloes are aligned with the cosmic web and how the kinematical nature of their web environment affects these alignments.

\subsection{Halo alignments with respect to the cosmic web}
\label{section:halo-shape}

 \begin{figure}
 \includegraphics[width=20pc]{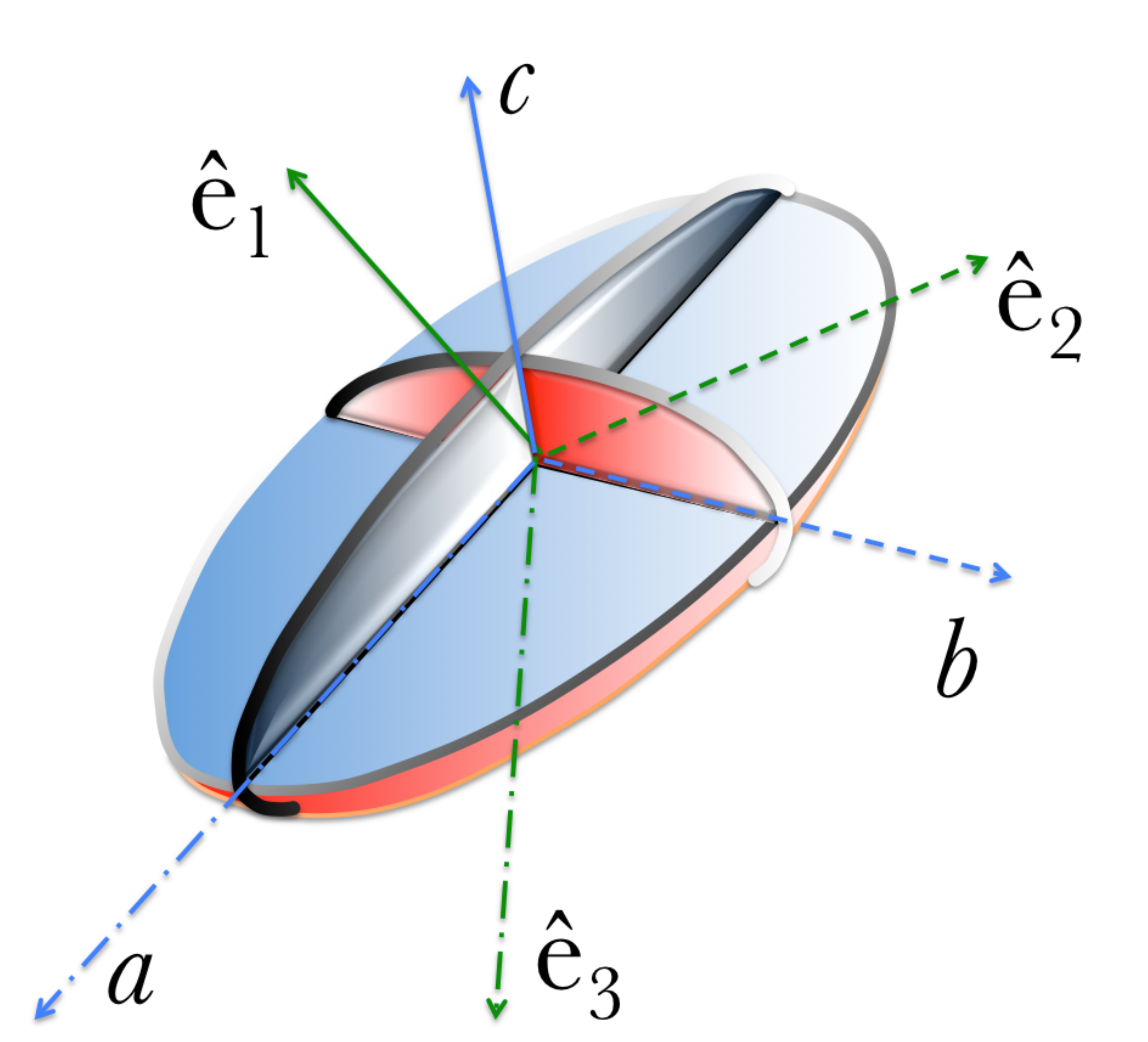}
 \caption{Schematic illustration of the angles examined in this work. The (reduced) inertia tensor is characterized by three principle axes ($\hat{a}$,~$\hat{b}$,~and~$\hat{c}$) that represent the long, intermediate and short axes of the halo respectively. The velocity shear is tensor is defined by the orthonormal basis $\hat{e}_{\rm 1}$, $\hat{e}_{\rm 2}$ and $\hat{e}_{\rm 3}$, where the eigenvalues and eigenvectors have been ordered to represent the axes along which material is collapsing fastest to slowest. We only consider the angles formed between vectors of the same line style, in other words $\hat{c}\cdot\hat{e}_{\rm 1}$, $\hat{b}\cdot\hat{e}_{\rm 2}$ and $\hat{a}\cdot\hat{e}_{\rm 3}$ }
 \label{fig:inertia}
 \end{figure}

\begin{figure}
 \includegraphics[width=20pc]{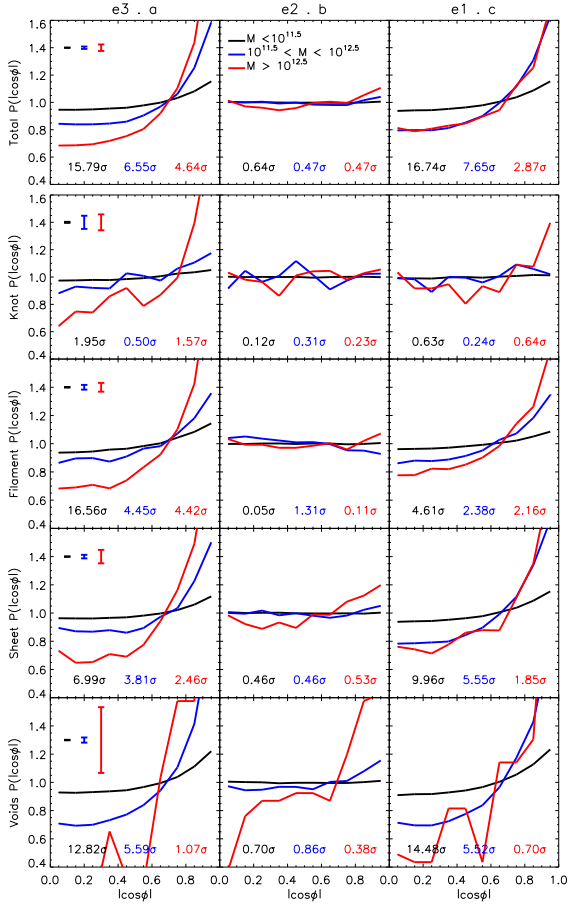}
\caption{The probability distribution $P(|\cos\phi|)$ as a function of the angle formed, $|\cos\phi|$, between the shape of a halo and the cosmic web. We show $\cos\phi=$\ethree$\cdot\hat{a}$ (left column), $\cos\phi=$\etwo$\cdot\hat{b}$ (middle column), and $\cos\phi=$\eone$\cdot\hat{c}$ (right column). The full angular distribution function (top row) is complemented by examining the alignment as a function of web environment, that is in haloes found in knots (second row), filaments (thtird row), sheets (fourth row) and voids (bottom row). Haloes are binned by mass, with low mass haloes (${\rm M}<10^{11.5}h^{-1}M_{\odot}$) in black, intermediate mass haloes ($10^{11.5}h^{-1}M_{\odot}<{\rm M}<10^{12.5}h{-1}M_{\odot}$) in blue and the most massive haloes (${\rm M}>10^{12.5}h^{-1}M_{\odot}$) in red. The strength of each distribution is characterized by the average offset between it and a random distribution in units of the Poisson error and is indicated by the colored numbers at the bottom of each panel. Note that even though the low mass bins often look ``flatter'', the signal is often more significant since the larger sample size decreases the Poisson error. The error bars in the left hand column indicate the expected poisson noise for a uniform distribution of that size.}
\label{fig:halo-shape}
 \end{figure}

 \begin{figure}
 \includegraphics[width=20pc]{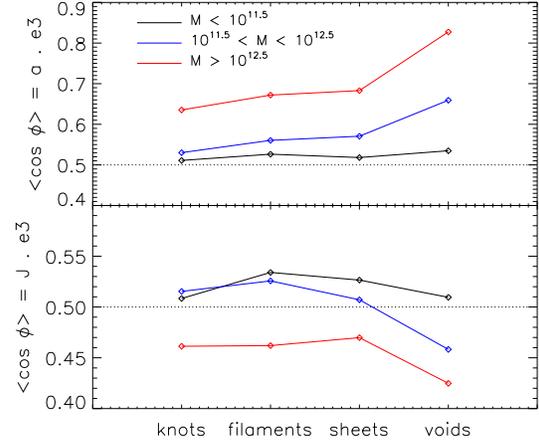}
 \caption{The median (cosine of the) angle formed between \ethree~ and the shape (top) and the spin (bottom) of haloes as a function of web type.The median angle for high, intermediate and low mass haloes are shown in red, blue and black respectively. A uniform distribution has a median of $<\cos\phi=0.5>$.}
 \label{fig:medians}
 \end{figure}

 \begin{figure}
 \includegraphics[width=20pc]{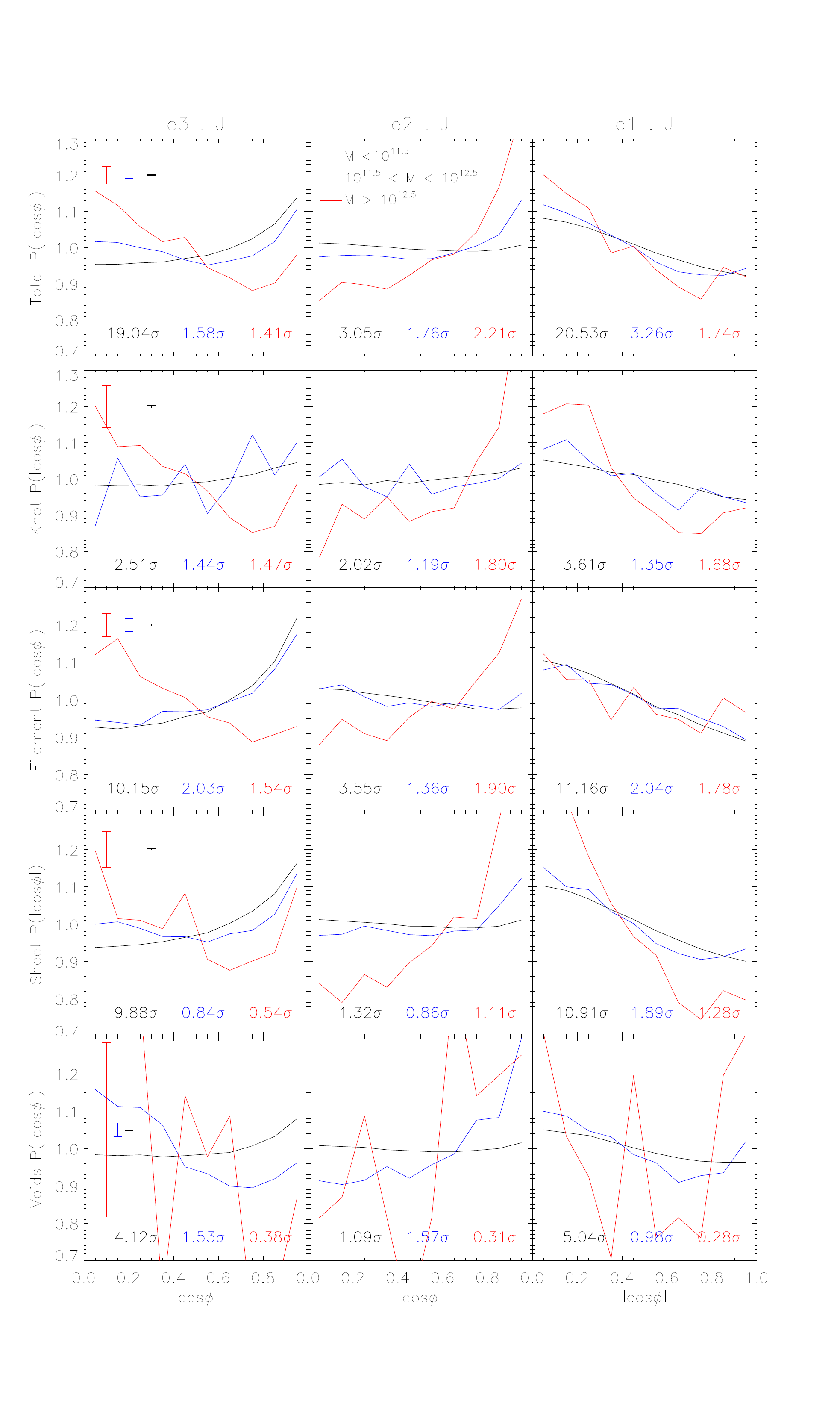}
 \caption{Same as Fig.~\ref{fig:halo-shape}, but for the angle formed between the halo's angular momentum vector $J$ and the eigenvectors of the shear tensor.}
 
 \label{fig:halo-spin}
 \end{figure}
 \begin{figure}
\includegraphics[width=20pc]{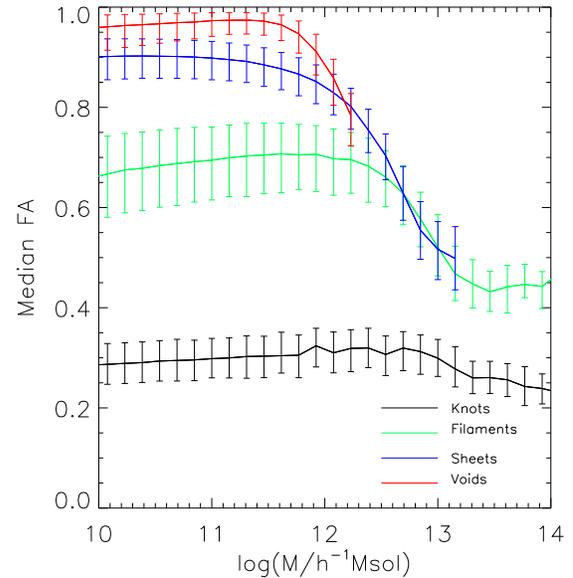}
 \caption{The median of the fractional anisotropy of the velocity shear in the local environment of a halo as a function of halo mass. We present the mass dependence of \textit{FA} for haloes found in knots, filaments, sheets and voids in black, green, blue and red respectively. Error bars represent the 1$\sigma$ spread about the median. Note that there exists a critical mass above which \textit{FA} decreases with increasing halo mass.}
 \label{fig:fa-fmass}
 \end{figure}

 \begin{figure}
 \begin{center}
 \includegraphics[width=20pc]{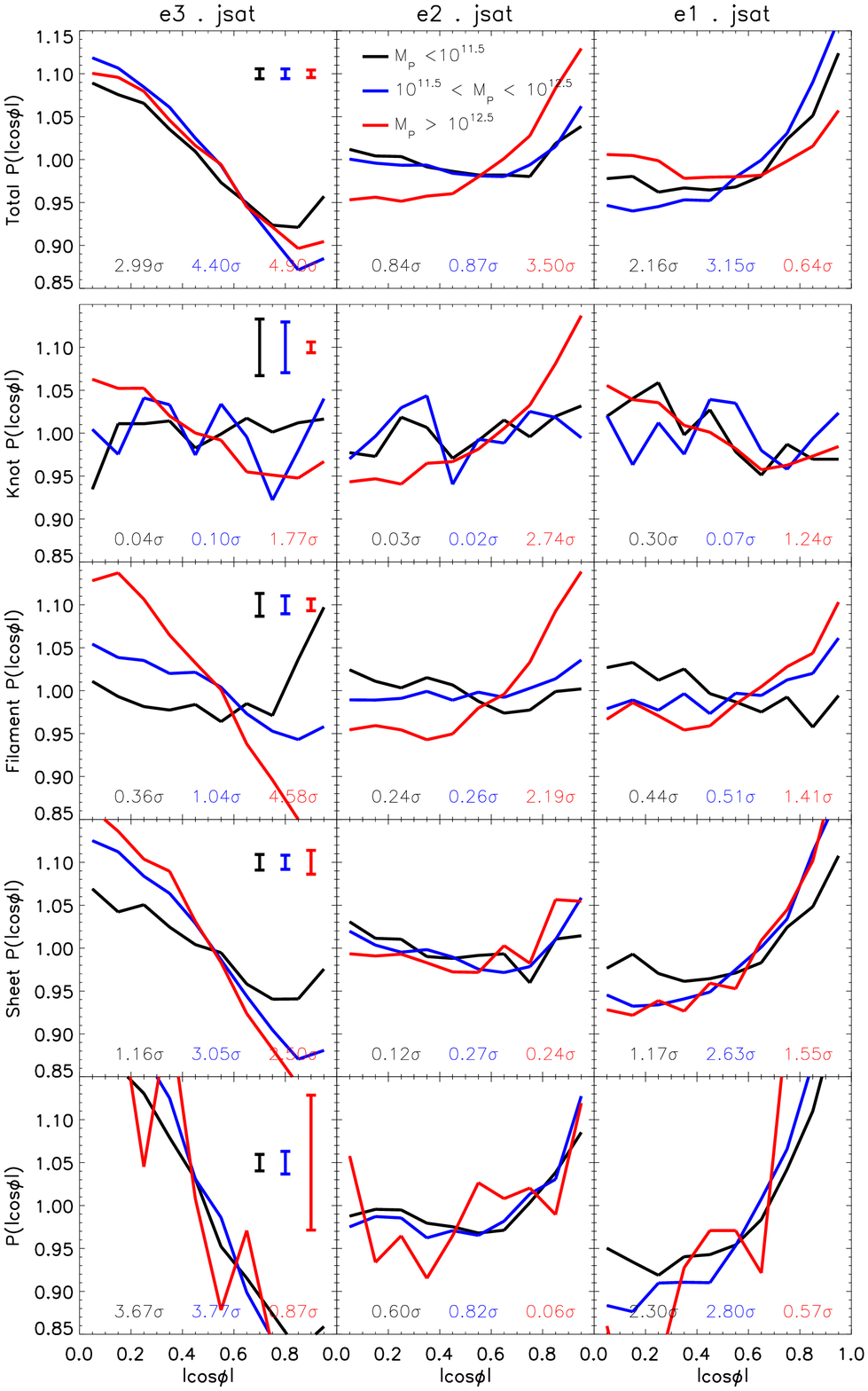}
 \caption{Same as Fig.~\ref{fig:halo-shape}, but for the angle formed between a  subhalo's orbital angular momentum vector $J$ and the eigenvectors of the shear tensor.}
 \label{fig:subhalo-spin}
\end{center}
 \end{figure}

The alignment  between haloes and the large scale structure can be quantified by examining the distribution of angles formed between each eigenvector of the diagonalized shear tensor and  a vector defining the halo. In what follows we investigate the alignment of the halo spin axis ($J$) and the halo shape ($\hat{a}$, $\hat{b}$, $\hat{c}$) with the local velocity shear. Since FOF haloes are not rotationally, these alignments are not identical.

When examining the alignment of haloes and subhaloes with the principle axes of the velocity shear tensor we choose to divide the halo sample by web element and by mass. The number of haloes in each mass bin and web type is presented in Table~\ref{table:halocounts}.

\subsubsection{Alignments with respect to halo shape}

In practice since there are two orthonormal bases (the velocity shear tensor and the inertia tensor) there are a total of 9 (=3$^{2}$) possible angles to examine. Since many of these are complimentary we choose to examine the angles formed between \eone~and $\hat{c}$, \etwo~and $\hat{b}$, and \ethree~and $\hat{a}$. These angles are chosen since if, e.g. a prolate halo points along the filament it lives in, $\cos\phi=\hat{e}_{3}.\hat{a}=0$. Similarly if a disc shaped halo lies in a sheet $\cos\phi=\hat{e}_{1}.\hat{c}=0$. Note that since the principle axes of both tensors have no direction (that is, they define an axis, not a direction), only the absolute value of the angle is considered.

In Fig.~\ref{fig:halo-shape} we show the alignment of halo shape with the cosmic web. In the upper row we show the alignment for the full population, while in the second, third, fourth and fifth row, we show the alignments in knots, filaments, sheets and voids respectively. From left to right the columns in Fig.~\ref{fig:halo-shape} show the distribution of angles formed between the principle axes, as mentioned above. 

The strength (or weakness) of the alignment signal can be quantified by $\sigma$, the average offset between a given distribution and a uniform one, calculated in terms of the Poisson error. In practice, the average difference between the number of haloes found in a given bin and the number expected from a uniform distribution is calculated in terms of the Poissonion error of a radnom distribution. If this is less than unity, then (on average) the distribution lies within the Poisson error of a uniform  distribution. High $\sigma$ indicates a strong deviation from uniformity while a weak signal corresponds to low $\sigma$. In each panel we indicate the strength (or weakness) of the alignment signal by a how far away it is from a uniform distribution. Generally speaking we say the vectors are aligned if the signal is stronger than $1\sigma$.

We begin by examining the full halo population irrespective of web classification. There are are two salient points to be inferred from the top row of Fig.~\ref{fig:halo-shape}.

\begin{enumerate}
\item \textit{Halos tend to point parallel to  \ethree.} While the distribution of the angle between \etwo~and $\hat{b}$ is consistent with random both $\cos\phi=\hat{e}_{3}.\hat{a}$ and $\cos\phi=\hat{e}_{1}.\hat{c}$ show statistically significant parallel alignments. This implies that long axis of a halo tends to align itself with \ethree.
\item \textit{The strength of the halo alignment is mass dependent}. Although the lowest mass haloes tend to show an alignment at just the $\sim$20\% level (for the angle between $\hat{e}_{3}$ and $\hat{a}$, and  $\hat{e}_{1}$ and $\hat{c}$), the more massive halo bins show considerably stronger alignments. This may reflect the tendency of more massive haloes to become more aspherical and for the weaker low mass alignment to be due to the ill-defined principle axes of low mass haloes \citep[i.e.][]{2006MNRAS.367.1781A}.
\end{enumerate}

Turning to the second, third, fourth and bottom rows of Fig.~\ref{fig:halo-shape} allows us to examine the web dependence of the shape alignment. Here too a mass dependence of the strength of the shape-alignment is seen for each web type. Surprisingly the alignment exists for all web types, implying that \textit{haloes are aligned with the shear tensor regardless of the web type they inhabit}. That is, the main tracer of the halo alignment appears to be the principle axes of the shearing material, and not whether a region is expanding or collapsing along any of the principle axes.

That said, the alignment for the low mass haloes appears to slightly increase in filaments and sheets with respect to knots. Presumably this is due to the increased environmental density in knots, which may have the effect of torquing low mass haloes and thus randomizing their alignments. The alignment of intermediate and high mass haloes on the other hand shows a strengthening in voids (although low number statistics may play a role). 

We quantify the effect of web environment on the halo shape alignment (with \ethree) in the top panel of Fig.~\ref{fig:medians} where the median of the angular distribution as a function of web classification is plotted for the three halo mass bins. A uniform distribution has a median of $<|\rm cos\phi|>=0.5$ (shown as the dotted line). The visual interpretation of the alignment distributions is confirmed - the median angle in all mass bins varies by at most $\sim5^{\circ}$ degrees for all mass bins in knots, filaments and sheets, and increases by $\sim15^{\circ}$ degrees for the highest mass haloes in voids. The slight strengthening of the median alignment for low mass haloes in filaments and sheets (by around $\sim 5^{\circ}$), is also visible.

On the right most panel in Fig.~\ref{fig:cosmicweb} we show graphically the orientation of halo shape with respect to the large scale structure in a thin slice through the simulation. Each halo is represented by an ellipse in the $x,y$-plane whose axes ratio correspond to those calculated by the inertia tensor. While the statistical nature of the alignment signal is visible, it is strong enough to pick out by eye that haloes point along the structure they are located in.

\begin{table*}
\begin{center}
 \begin{tabular}{r r r r r}
  & $M < 10^{11.5}$ & $10^{11.5}< M < 10^{12.5}$  &$M > 10^{12.5}$ & all Masses\\
   \hline
   \hline
   \hline
Total  & 8,742,899 & 132,658 & 17,771 &  8,893,328 \\
   \hline
Knots  &          8\%   &     3\% &    17\% & 8\%\\
Filaments & 23\%  &  27\%   & 57\%  & 23\%\\
Sheets  &     35\%  &  47\%   &   25\%&  35\%\\
Voids &         34\% &  23\%    &   1\% &  34\%\\
    \hline

 \end{tabular}
 \end{center}
\caption{The number of haloes in each mass bin and the fraction thereof in each web type, used in this study. Masses are given in units of $h^{-1}M_{\odot}$. }
\label{table:halocounts}
\end{table*}

\subsubsection{Alignments with respect to halo spin}

We now turn to the (kinematical) alignment of halo spin with the shear tensor. Here too, the absolute value of the angle formed between $J$ and $\hat{e}_{i}$ is presented since $\hat{e}_{i}$ does not denote a proper geometric direction.

In Fig.~\ref{fig:halo-spin} we show the probability distribution of the angle formed between the angular momentum vector and \ethree, \etwo~ and \eone~for all web types as well as for each individually. Let us begin be highlighting the similarities as found with the shape alignment in Sec.~\ref{section:halo-shape} and Fig.~\ref{fig:halo-shape}.

The alignment between spin and the cosmic web is significantly weaker than that found using the shape in the previous section. As in the shape-alignment, the parallel alignment of spin with \etwo~and perpendicular alignment with \eone~is web independent. Each web type (noisiness due to different sample size not withstanding) tells more or less the same story: the orientation and strength of halo spin alignment is mass dependent.

The alignment between $J$ and \ethree, is fairly well nuanced. As shown in previous work \citep[e.g.][]{2012arXiv1201.5794C,Aragon-Calvoetal2007,2012arXiv1201.6108T} high mass haloes display perpendicular alignments with \ethree, while lower mass haloes show a parallel alignment. Note however that in previous work \ethree~was defined specifically in the case of filaments or walls - here we show that the flipping of high mass halo alignment occurs with respect to \ethree~\textit{regardless} of web type. The mass at which the transition occurs \citep[defined by][as $5\times10^{12}M_{\odot}$]{2012arXiv1201.5794C} appears to decrease with web type, such that halos in the intermediate mass bin show a spin-alignment flip in voids. This finding suggests that spin alignments are qualitatively web independent but the strengths of these alignments are  web dependent.

The alignment of the halo spin axis with the other two eigenvectors (\eone~and \etwo) behaves more similarly with respect to mass and web type. In general, that spins tend to align with \etwo~was first suggested by \cite{NavarroAbadiSteinmetz04}. Again, this alignment is strengthened for the most massive halo bin. Halo spin axes also tend to be perpendicular to \eone, the direction of the greatest collapse (in knots, filaments and sheets) and of the weakest expansion in voids.

As in the previous sub section, we quantify the strength of the alignment by looking at the median of each distribution in the bottom panel of Fig.~\ref{fig:medians}. A similar picture to that presented with respect to the long axis is seen, namely that the median alignment is roughly constant (changing by less than $\sim5^{\circ}$) for knots, filaments and sheets, and strengthens in voids for the two highest mass bins. That the high mass bin always has a median less than 0.5 reflects the tendency for the most massive haloes to orient themselves away from \ethree.

\subsection{The effect of shear anisotropy on halo mass}

In the preceding section it was shown that alignments between halo shape and the cosmic web appear dependent on mass and independent on web type (with the exception of voids). Although the halo spin-web alignment has a transitional mass above which the alignment tends towards perpendicular, this mass appears web dependent decreasing with decreasing web type (that is, higher in knots and lower in voids). As was already shown in Fig.~\ref{fig:fa-distr}, each web type has a fairly different distribution of web sphericity, with voids defining the most aspherical structures while knots collapse more uniformly. Given that the transitional mass in the halo-spin alignment appears web dependent, it is interesting to ask if the \textit{FA}, as measured by equation~\ref{eq:fa}, of the velocity shear plays a role in the determination of halo mass. Or to put it another way: if the most massive haloes display the strongest alignment, is their environment (as defined by shear \textit{FA}) responsible?

In Fig.~\ref{fig:fa-fmass} we present the median shear \textit{FA} as a function of halo mass for each halo subdivided into its web type. A number of interesting observations can be made. First, low mass haloes show no (or only a very weak) dependence on shear \textit{FA}. At a specific halo mass however, the flat relationship disappears and an inverse relation between halo mass and shear \textit{FA} is found. The most massive haloes exist in regions of near uniform collapse. Insight into halo assembly may be thus gained by this relation: a halo may only grow to large masses if the region it is in is uniformly collapsing. Large shears inhibit halo growth. On the other hand, smaller haloes can be born in a myriad of shear environments.

This behavior is qualitatively web independent. However it differs quantitatively when examining the relation for the four different web types. In a given web type, the mass at which the \textit{FA} begins to become mass dependent differs - lower for voids, greater for filaments, sheets and knots. Although it is difficult to quantify the mass at which \textit{FA} becomes mass dependent, if one examines Fig.~\ref{fig:fa-fmass} by eye, it appears to be similar to the transitional mass for the flipping of spin-alignment. It is thus implied that the \textit{FA} of the halo's shearing environment determines halo mass and spin in a statistical sense.

\subsection{Subhalo alignments with respect to the cosmic web}

Since the web finder employed in this work probes the non-linear regions of the large scale structure, it is interesting to examine whether the positions or orbits of subhaloes deep within their parent potential are affected by the large scale structure. To do so, it is insightful to examine similar plots as shown in Fig.~\ref{fig:halo-shape} and Fig.~\ref{fig:halo-spin} but for subhalos, not host haloes. 
Although examining the spatial distribution of subhaloes with respect to the LSS would be indeed very revealing \citep[e.g.][]{Libeskindetal2005,Libeskindetal2007,Libeskindetal2009,MetzKroupaLibeskind2008,MetzKroupaJerjen2009,KroupaTheisBoily2005,2012arXiv1206.1340W}, by looking at the ensemble of subhalo orbital angular momentum, spatial information is implicitely included\footnote{This is simply because $J_{\rm sat}$ is, by definition, perpendicular to $r_{\rm sat}$, the satellites position vector, and thus the distribution of $J_{\rm sat}$ reflects the distribution of $r_{\rm sat}$.}.

In Fig.~\ref{fig:subhalo-spin} we present the distribution of the cosine of the angle formed between the orbital angular momentum of satellites and the three principle axes of the shear tensor. As in Fig.~\ref{fig:halo-shape} and Fig.~\ref{fig:halo-spin}, we present the full distribution in the top row, and the alignment signal in knots, filaments, sheets and voids in the second, third, fourth and fifth row.

The subhalo orbital alignment behaves differently than the halo spin alignment. Note that low mass host haloes will not contribute any subhaloes to this sample while massive host haloes will contribute many. We begin by examining the full signal irrespective of web type. 

Subhalo orbital angular momentum, denoted as $J_{\rm sat}=r_{\rm sat} \times v_{\rm sat}$, show a weaker (at the $\sim10\%$ level) but still statistically significant parallel alignment with \eone~and perpendicular alignment with \ethree. Since subhalos tend to orbit along (or in the plane of) their parent's major axis, and since the parent major axis tends to be aligned with \ethree, it follows that the sub halo orbits should tend to be perpendicular to \ethree. Note that the flipping of subhalo orbital spin alignment is not seen for subhaloes when binned by parent mass.

The strength of the alignment signal and its sense is quite clearly web and mass dependent. Fig.~\ref{fig:subhalo-spin} shows that the signal for a given parent halo mass bin gets weaker with web type from void to knot. Indeed the greatest mass bin shows a correlation with \ethree~for all web types: $\sim 7\%$ in knots growing to $\sim 15\%$ in voids). For the intermediate mass bin a correlation only emerges in filaments, sheets and voids while in the lowest mass bin it is only seen in sheets and voids. The weakening of subhalo orbital spin alignment in knots is likely due to the denser environment where the abundance of neighboring haloes provides torques capable of weakening the correlation. Finally we note that in knots, the subhalo angular momentum alignment with \eone~is actually opposite in its sense to that in the other web types. 

In summary, the web environment and mass of each host halo has a direct impact on the alignment of its subhaloes. While only the most massive parents display an alignment of $j_{\rm sat}$ with \ethree~in knots, even the least massive parent show the same alignment in voids. A similar picture is seen with respect to \eone~and \etwo.

\section{Summary and Discussion}
\label{sec:conclusion}

The shape and spin of DM haloes and the angular momentum of their subhaloes define characteristic directions. These orientations are studied within the context of the large scale structure (LSS). The LSS exhibits a web-like structure, the so-called cosmic web, composed of knots, filaments, sheets and voids. These are characterized here by the velocity shear of the cosmic fluid. It is this velocity shear, engendered by the primordial perturbation field, which is believed to have endowed galaxies with their angular momentum through the process of tidal torquing. The main focus of this work is to study the alignment of the characteristic directions of each halo with those of the LSS, in order to attempt to understand the alignment of haloes with the LSS.

The eigenvectors of the velocity shear tensor provide a common framework within which the directional properties of DM halos and of the LSS can be studied and compared \citep{2012arXiv1201.3367H,2012MNRAS.421L.137L}. The results presented here are based on the study of the Bolshoi simulation (Klypin et al. 2011). The wide dynamical range and resolution of the simulation has enabled us to penetrate deeper into the non-linear regions surrounding haloes at the present time.

DM halos and the cosmic web constitute two different manifestations of the dynamical evolution of the LSS. Both are scale dependent and have their specific directional properties. Any attempt to compare these two should start with selecting an effective dynamical resolution for both halos and the cosmic web. While a very coarse grid runs the risk of washing out any alignment signal (due to a Gaussian smoothing of the velocity field that is applied), a fine grid will probe the inner depths of each halo and not the large scale structure. Thus our selection of a grid whose cell size if  $\sim$1Mpc$/h$, puts an upper limit of $\sim2\times10^{14}M_{\odot}/h$ on the halo size we may probe.

The eigenvectors of the velocity shear tensor provide an orthonormal basis of unit vectors that define the principal directions of the cosmic web.  We use this basis to investigate the alignment of halo spin, shape (as defined by its moment of inertia) and subhalo orbital angular momentum with the respect to the LSS. While much previous work has been dedicated to examining these alignments, nearly all the literature focuses on just one or two web types individually since most cosmic web finders are tailor made for one specific web type (e.g. filament or void finders). To the best of our knowledge, we are thus the first paper to examine halo alignments with respect to the velocity shear tensor, and in particular to all cosmic web elements simultaneously.

Using this democratic approach, we have found a strong alignment of halo shape with the cosmic web. This is a confirmation of  number of studies \citep[e.g.][]{2006MNRAS.370.1422A,Aragon-Calvoetal2007,2007MNRAS.375..184B,2009ApJ...706..747Z} which looked at shape alignment of haloes in one or two individual web types. The alignment signal found here is statistically significant across the entire cosmic web and can be seen graphically in the right most panel on Fig.~\ref{fig:cosmicweb}. This forms the bulk of one of our main results: \textit{the alignment between halo shape and the cosmic web is web independent: the shear tensor traces this alignment}. The alignment is visible for all web types when using the velocity shear to quantify the cosmic web. From this correlation, a conjecture may be made: the velocity shear is the sole determining factor of halo alignment. Further work at higher redshift is necessary to verify this conjecture. We defer this calculation to a future paper. Yet we still may generalize previous explanations which are limited by web type (i.e. accretion along a filament, etc)  and suggest that the velocity shear is the more fundamental tracer of web independent halo alignments.

The alignment of halo spin with the cosmic web has thus been examined in the context of previous findings which indicated that lower mass haloes tend to spin with their rotation axis aligned parallel to the large scale structure in filaments and walls. The alignment of high mass haloes in filaments on the other hand was found to tend to be perpendicular to the filament's axis. The mass where this flip occurs was found in previous studies to be $\sim5\times10^{12}M_{\odot}$ \citep{Aragon-Calvoetal2007}.  Studies such as \cite{2012arXiv1201.5794C} attributed this flip in the orientation of the rotation axis to the merger history - whereas low mass haloes grow by isotropic accretion of diffuse dark matter, higher mass haloes grow by the merger of subhaloes \citep[e.g.][]{2009MNRAS.399..762F} whose infall is correlated with the large scale structure \citep{Knebeetal2004,Zentneretal2005,Libeskindetal2005,Libeskindetal2011a}.  \cite{2012MNRAS.tmpL.511L} have suggested that the orientation of spins may be due to the assembly bias of high mass halos - a high mass halo dominates the accretion and hence the angular momentum in its local neighborhood at the expense of low mass haloes. Most of these studies implicitly assign the mass scale at which halo growth becomes merger dominated as the transitional spin flip mass of $\sim5\times10^{12}M_{\odot}$.  \cite{2012arXiv1201.6108T} have argued against this interpretation as they find that the spin axes of all haloes are oriented perpendicular to filaments at high $z$, and it is just the spin axes of low mass haloes that evolve towards a parallel alignment.

While considerably weaker than the shape alignment, we confirm the results of other studies that focus on the spin alignment. Yet we find that the mass at which the spin flips to be web dependent, hinting that instead of reflecting a physical process associated with the halo it is more likely a reflection of the interaction with the web environment and thus the scale against which it is measured. 
We have found that (at a fixed scale of $\sim$1Mpc$/h$) the transitional mass discriminating between perpendicular or parallel alignment decreases from $\sim10^{12.5}h^{-1}M_{\odot}$ in knots to $\sim10^{11.5}h^{-1}M_{\odot}$ in voids. Furthermore the mass at which the spin flips with respect to the large scale structure appears to be intimately related to the mass at which haloes become dependent on the shear's uniformity (see below).

Since haloes are assigned the eigenvalues of the shearing material they are embedded in, we have compared this distribution with the full distribution of all cell eigenvalues. These distributions differ indicating a significant bias for studies that attempt to reconstruct the large scale shear based on magnitude limited samples \citep[e.g.][]{LeeErdogdu2007}.  Just as galaxies represent a biased sample of the density field (they trace high density peaks), so too is the shear as measured by haloes a biased representation.

The shear uniformity has been characterized using the fractional anisotropy \textit{FA} \citep[][see footnote 2]{Basser1995} which measures how much of the shear magnitude is in an anisotropic element. A strong asymmetry in the distribution is found, in the sense that knots appear to collapse more isotropically than voids expand (the collapse of a knot is not the expansion of a void played in reverse). Halo mass appears to correlate with $FA$ above a certain mass: more massive haloes inhabit regions where the shear is more spherical. The mass at which this dependence occurs is roughly the mass scale for which spin-alignments flip.

We have confirmed that the alignment of subhalo orbits is also well correlated with the large scale structure, albeit unlike the haloes the correlation is web dependent. Possibly owing to the denser environment in knots, the correlation is stronger in voids and weaker in knots. The large scale structure as defined by the velocity shear thus has a direct influence on the orbits of subhaloes deep in their parent potentials. This finding has a direct bearing on the correlated rotation of the Milky Way's own satellite system \citep{MetzKroupaJerjen2009}.

As future surveys provide more and more detailed photometry of galaxies in the framework of cosmic web, we will undoubtedly be able to infer more and more regarding the importance of environment as defined dynamically on the process of dm halo assembly and galaxy formation. Observations of the spatial distribution of external satellite galaxies as well as the orientation of the spins of extra-galactic discs and spheroids with respect to the cosmic web, will help us understand the importance of the web in determining how galaxies acquire their angular momentum.

\section*{Acknowledgments}
This work was supported by the Deutsche Forschungs Gemeinschaft, the Deutsche Akademische Austausch Dienst, the Ministerio de Ciencia e Innovacion in Spain, the Ramon y Cajal program, (AYA 2009-13875-C03-02, AYA2009-12792-C03-03, CAM S2009/ESP-1496, FPA2009-08958) and through Consolider-Ingenio  SyeC. This research was also supported in part by the National Science Foundation under Grant No. NSF PHY11-25915. The Bolshoi simulation and analysis was run on the Pleiades supercomputer at NASA Ames Research Center. YH has been partially supported by the Israel Science Foundation (13/08).
\bibliography{./ref}
\end{document}